\documentclass[12 pt,twoside,prd,amsmath,amssymb,tightenlines,showpacs,
showkeys,checkins,eqsecnum]{revtex4}
\usepackage{epsfig}
\usepackage{psfig}
\usepackage{mathptmx}
\usepackage{bm}
\usepackage{graphicx}

\usepackage{hyperref}
\date{\filedate}

\renewcommand{\cal}{\mathcal}

\newcommand {\ve}{\varepsilon}

\def \myfigures #1#2#3#4#5#6#7#8
{\begin{figure}[ht]
    \begin{center}
        \includegraphics[width=#2 \textwidth]{#1.eps}
        \hfill
        \includegraphics[width=#6 \textwidth]{#5.eps}
        \parbox[t]{#4\textwidth}{\caption {#3}\label{#1}}
        \hfill
        \parbox[t]{#8\textwidth}{\caption {#7}\label{#5}}
    \end{center}
\end{figure}}

\def \myfigs #1#2#3#4#5#6#7#8
{\begin{figure}[ht]
    \begin{center}
        \includegraphics[width=#2 \textwidth, angle=-90]{#1.eps}
        \hfill
        \includegraphics[width=#6 \textwidth, angle=-90]{#5.eps}
        \vskip 5 mm
        \parbox[t]{#4\textwidth}{\caption {#3}\label{#1}}
        \hfill
        \parbox[t]{#8\textwidth}{\caption {#7}\label{#5}}
    \end{center}
\end{figure}}

\def\myfigure #1#2#3#4
{\begin{figure}[ht]\begin{center}
\includegraphics[width=#2 \textwidth]{#1.eps}
\parbox[t]{#4\textwidth}{\caption{#3}\label{#1}}
\end{center}\end{figure}}

\def\myfig #1#2#3#4
{\begin{figure}[ht]\begin{center}
\includegraphics[width=#2 \textwidth, angle = -90]{#1.eps}
\vskip 1 mm
\parbox[t]{#4\textwidth}{\caption{#3}\label{#1}}
\end{center}
\end{figure}}

\begin{document}
\title{Anisotropic cosmological models with perfect fluid and dark energy
revisited}
\author{Bijan \surname{Saha}}
\affiliation{Laboratory of Information Technologies\\
Joint Institute for Nuclear Research, Dubna\\
141980 Dubna, Moscow region, Russia} \email{saha@thsun1.jinr.ru, bijan@jinr.ru}
\homepage{http://thsun1.jinr.ru/~saha/}
\date{\today}
\begin{abstract}
We consider a self-consistent system of Bianchi type-I (BI)
gravitational field and a binary mixture of perfect fluid and dark
energy. The perfect fluid is taken to be the one obeying the usual
equation of state, i.e., $p = \zeta \ve$, with $\zeta \in [0,\,1]$
whereas, the dark energy is considered to be obeying a
quintessence-like equation of state. Exact solutions to the
corresponding Einstein equations are obtained. The model in
consideration gives rise to a Universe which is spatially finite.
Depending on the choice of problem parameters the Universe is
either close with a space-time singularity, or an open one which
is oscillatory, regular and infinite in time.
\end{abstract}



\maketitle

\bigskip

            \section{Introduction}

The discovery that the expansion of the Universe is accelerating
\cite{Bachall} has promoted the search for new types of matter
that can behave like a cosmological constant
\cite{caldwell,starobinsky} by combining positive energy density
and negative pressure. This type of matter is often called {\it
quintessence}. Zlatev {\it et al.} \cite{zlatev} showed that
"tracker field", a form of qiuntessence, may explain the
coincidence, adding new motivation for the quintessence scenario.

An alternative model for the dark energy density was used by
Kamenshchik {\it et al.} \cite{kamen}, where the authors suggested
the use of some perfect fluid but obeying "exotic" equation of
state. This type of matter is known as {\it Chaplygin gas}. In
doing so the authors considered mainly a spatially flat,
homogeneous and isotropic Universe described by a
Friedmann-Robertson-Walker (FRW) metric.

The theoretical arguments and recent experimental data, which
support the existence of an anisotropic phase that approaches an
isotropic one, lead to consider the models of Universe with
anisotropic back-ground. Since the modern-day Universe is almost
isotropic at large, its simplicity and evolution into a FRW
Universe makes the BI Universe a prime candidate for studying the
possible effects of an anisotropy in the early Universe on
present-day observations. In a number of papers, e.g.,
\cite{PRD23501,PRD24010}, we have studied the role of a nonlinear
spinor and/or a scalar fields in the formation of an anisotropic
Universe free from initial singularity. It was shown that for a
suitable choice of nonlinearity and the sign of $\Lambda$ term the
model in question allows regular solutions and the Universe
becomes isotropic in the process of evolution. Recently
Khalatnikov {\it et al.} \cite{khalat} studied the Einstein
equations for a BI Universe in the presence of dust, stiff matter
and cosmological constant. In a recent paper \cite{lambda} the
author studied a self-consistent system of Bianchi type-I (BI)
gravitational field and a binary mixture of perfect fluid and dark
energy given by a cosmological constant. The perfect fluid in that
paper was chosen to be the one obeying either the usual equation
of state, i.e., $p = \zeta \ve$, with $\zeta \in [0,\,1]$ or a van
der Waals equation of state. That paper was followed by another
where we studied the evolution of an initially anisotropic
Universe given by a BI spacetime and a bimnary mixture of a
perfect fluid obeying the equation of state $p = \zeta \ve$ and a
dark energy given by either a quintessence or a Chaplygin gas
\cite{denpf}. It should be mentioned that the inclusion of dark
energy does not eliminate initial singularity of the model and the
space-time in those cases is ever-expanding. In order to obtain a
singularity-free Universe in the present paper we introduce a
modified version of quintessence-like dark energy, which at the
same time is able to explain the accelerated expansion.

            \section{Basic equations}
The gravitational field in our
case is given by a Bianchi type I (BI) metric in the form
\begin{equation}
ds^2 =  dt^2 - a^2 dx^2 - b^2 dy^2 - c^2 dz^2,
\label{BI}
\end{equation}
with the metric functions $a,\,b,\,c$ being the functions of time
$t$ only.

The Einstein field equations for the BI space-time we write in the form
\begin{subequations}
\label{ee}
\begin{eqnarray}
\frac{\ddot b}{b} +\frac{\ddot c}{c} + \frac{\dot b}{b}\frac{\dot
c}{c}&=&  \kappa T_{1}^{1},\label{11}\\
\frac{\ddot c}{c} +\frac{\ddot a}{a} + \frac{\dot c}{c}\frac{\dot
a}{a}&=&  \kappa T_{2}^{2},\label{22}\\
\frac{\ddot a}{a} +\frac{\ddot b}{b} + \frac{\dot a}{a}\frac{\dot
b}{b}&=&  \kappa T_{3}^{3},\label{33}\\
\frac{\dot a}{a}\frac{\dot b}{b} +\frac{\dot b}{b}\frac{\dot c}{c}
+\frac{\dot c}{c}\frac{\dot a}{a}&=&  \kappa T_{0}^{0}.
\label{00}
\end{eqnarray}
\end{subequations}
Here $\kappa$ is the Einstein gravitational constant and over-dot means
differentiation with respect to $t$. The energy-momentum tensor of the
source is given by
\begin{equation}
T_{\mu}^{\nu} = (\ve + p) u_\mu u^\nu - p \delta_\mu^\nu,
\label{emt}
\end{equation}
where $u^\mu$ is the flow vector satisfying
\begin{equation}
g_{\mu\nu} u^\mu u^\nu = 1.
\label{scprod}
\end{equation}
Here $\ve$ is the total energy density of a perfect fluid and/or
dark energy density, while $p$ is the corresponding pressure. $p$
and $\ve$ are related by an equation of state which will be
studied below in detail. In a co-moving system of coordinates from
\eqref{emt} one finds
\begin{equation}
T_0^0 = \ve, \qquad T_1^1 = T_2^2 = T_3^3 = - p.
\label{compemt}
\end{equation}
In view of \eqref{compemt} from \eqref{ee} one immediately obtains
\cite{PRD23501}
\begin{subequations}
\label{abc}
\begin{eqnarray}
a(t) &=&
D_{1} \tau^{1/3} \exp \bigl[X_1 \int\,\frac{dt}{\tau (t)} \bigr],
\label{a} \\
b(t) &=&
D_{2} \tau^{1/3} \exp \bigl[X_2 \int\,\frac{dt}{\tau (t)} \bigr],
\label{b}\\
c(t) &=&
D_{3} \tau^{1/3}\exp \bigl[X_3  \int\,\frac{dt}{\tau (t)} \bigr].
\label{c}
\end{eqnarray}
\end{subequations}
Here $D_i$ and $X_i$ are some arbitrary constants obeying
$$D_1 D_2 D_3 = 1, \qquad X_1 + X_2 + X_3 = 0,$$
and $\tau$ is a function of $t$ defined to be
\begin{equation}
\tau = a b c. \label{tau}
\end{equation}
Summation of \eqref{11}, \eqref{22}, \eqref{33} and 3 times
\eqref{00} gives the equation for $\tau$ :
\begin{equation}
\frac{\ddot \tau}{\tau} = \frac{3 \kappa}{2} \bigl(\ve - p\bigr),
\label{dtau}
\end{equation}
whereas, from the conservation law for the energy-momentum tensor
for $\ve$ we find
\begin{equation}
\dot{\ve} = -\frac{\dot \tau}{\tau} \bigl(\ve + p\bigr).
\label{dve}
\end{equation}
After a little manipulations from \eqref{dtau} and \eqref{dve} we
find
\begin{equation}
\dot \tau = \pm \sqrt{C_1 + 3 \kappa \ve \tau^2}, \label{1st}
\end{equation}
with $C_1$ being an integration constant. Eq. \eqref{dve} can be
rewritten in the form
\begin{equation}
\frac{\dot{\ve}}{(\ve + p)} = -\frac{\dot \tau}{\tau}.
\label{dve1}
\end{equation}
Taking into account that the pressure and the energy density obey
a equation of state of type $p = f (\ve)$, we conclude that $\ve$
and $p$, hence the right hand side of the Eq. \eqref{dtau} is a
function of $\tau$ only, i.e.,
\begin{equation}
\ddot \tau = \frac{3 \kappa}{2} \bigl(\ve - p\bigr)\tau \equiv
{\cal F} (\tau). \label{dtau1}
\end{equation}
From the mechanical point of view Eq. \eqref{dtau1} can be
interpreted as an equation of motion of a single particle with
unit mass under the force ${\cal F}(\tau)$. Then the following
first integral exists \cite{landau}:
\begin{equation}
\dot \tau = \sqrt{2[{\cal E} - {\cal U} (\tau)]}. \label{first}
\end{equation}
Here ${\cal E}$ can be viewed as energy and ${\cal U}(\tau)$ is
the potential of the force ${\cal F}$. Comparing the Eqs.
\eqref{1st} and \eqref{first} one finds ${\cal E} = C_1 /2$ and
\begin{equation}
{\cal U} (\tau) = - \frac{3}{2} \kappa \ve \tau^2.
\label{potential}
\end{equation}

Finally, rearranging \eqref{1st}, we write the solution to the Eq.
\eqref{dtau} in quadrature:
\begin{equation}
\int \frac{d\tau}{\sqrt{C_1 + 3\kappa \ve \tau^2}} = t + t_0,
\label{quad}
\end{equation}
where the integration constant $t_0$ can be taken to be zero,
since it only gives a shift in time.

In what follows we study the Eqs. \eqref{dtau} and \eqref{dve} for
perfect fluid and/or dark energy for different equations of state
obeyed by the source fields.

\section{Universe as a binary mixture of perfect fluid and dark energy}

In this section we thoroughly study the evolution of the BI
Universe filled with perfect fluid and dark energy in details.
Taking into account that the energy density ($\ve$) and pressure
($p$) in this case comprise those of perfect fluid and dark
energy, i.e.,
$$\ve = \ve_{\rm pf} + \ve_{\rm DE}, \qquad p = p_{\rm pf} + p_{\rm
DE}$$ the energy momentum tensor can be decomposed as
\begin{equation}
T_{\mu}^{\nu} = (\ve_{\rm DE} + \ve_{\rm pf} + p_{\rm DE} + p_{\rm
pf} ) u_\mu u^\nu - (p_{\rm DE} + p_{\rm pf}) \delta_\mu^\nu.
\label{emtde}
\end{equation}
In the above equation $\ve_{\rm DE}$ is the dark energy density,
$p_{\rm DE}$ its pressure. We also use the notations $\ve_{\rm
pf}$ and $p_{\rm pf}$ to denote the energy density and the
pressure of the perfect fluid, respectively.

In a comoving  frame the conservation law of the energy momentum
tensor leads to the balance equation for the energy density
\begin{equation}
\dot{\ve}_{\rm DE} + \dot{\ve}_{\rm pf} = - \frac{\dot \tau}{\tau}
\bigl(\ve_{\rm DE} + \ve_{\rm pf} + p_{\rm DE} + p_{\rm pf}
\bigr). \label{dveden}
\end{equation}
The dark energy is supposed to interact with itself only and it is
minimally coupled to the gravitational field. As a result the
evolution equation for the energy density decouples from that of
the perfect fluid, and from  Eq. \eqref{dveden} we obtain two
balance equations
\begin{subequations}
\label{balance}
\begin{eqnarray}
\dot{\ve}_{\rm DE} + \frac{\dot \tau}{\tau} \bigl(\ve_{\rm DE} +
p_{\rm DE}\bigr) &=& 0, \label{deve}\\ \dot{\ve}_{\rm pf} +
\frac{\dot \tau}{\tau} \bigl(\ve_{\rm pf} + p_{\rm pf}\bigr) &=&
0. \label{pfve}
\end{eqnarray}
\end{subequations}

\subsection{Equations of state}

In order to complete the system of equations we need to specify
two equations of state for $p_{\rm pf}$ and $p_{\rm DE}$.

\subsubsection{perfect fluid}

There are some equation of states that are commonly used that,
although not widely applicable, are obtained as a result of
approximate estimates for particular fluid. The barotropic
equation of state
\begin{equation}
p_{\rm pf}\,=\,\zeta\,\ve_{\rm pf}, \label{eqst}
\end{equation}
is often assumed. Here $\zeta$ is a constant and lies in the
interval $\zeta\, \in [0,\,1]$. Note that $0 \le \zeta \le 1$ is
necessary for the existence of local mechanical stability and for
the speed of sound in the fluid to be no greater than the speed of
light. Depending on its numerical value, $\zeta$ describes the
following types of Universes \cite{jacobs}
\begin{subequations}
\label{zeta}
\begin{eqnarray}
\zeta &=& 0, \qquad \qquad {\rm (dust\,\, Universe)},\\
\zeta &=& 1/3, \quad \qquad {\rm (radiation\,\, Universe)},\\
\zeta &\in& (1/3,\,1), \quad {\rm (hard\,\, Universes)},\\
\zeta &=& 1, \quad \qquad \quad {\rm (Zel'dovich\,\, Universe \,\,
or\,\, stiff\,\, matter)}.
\end{eqnarray}
\end{subequations}

In view of the Eq. \eqref{eqst} from \eqref{pfve} one easily finds
laws of change of energy density and pressure of a perfect fluid
with the expansion of the Universe :
\begin{equation}
\ve_{\rm pf} = \ve_0/\tau^{(1+\zeta)}, \qquad p_{\rm pf} = \ve_0
\zeta/\tau^{(1+\zeta)},
 \label{vepf}
\end{equation}
where $\ve_0$ is the integration constants. In absence of the dark
energy one immediately finds
\begin{equation}
\tau = C t^{2/(1+\zeta)}, \label{tpf}
\end{equation}
with $C$ being some integration constant. As one sees from
\eqref{abc}, in absence of a dark energy, for $\zeta < 1$ the
initially anisotropic Universe eventually evolves into an
isotropic FRW one, whereas, for $\zeta = 1$, i.e., in case of
stiff matter the isotropization does not take place.

\subsubsection{Dark energy}

It was mentioned earlier that the dark energy can be given by a
$\Lambda$ term, a quintessence or a Chaplygin gas. It was also
mentioned that the quintessence was constructed by combining
positive energy density and negative pressure and obeys the
equation of state
\begin{equation}
p_{\rm q} = w_{\rm q} \ve_{\rm q}, \label{quint}
\end{equation}
where the constant $w_{\rm q}$ varies between $-1$ and zero, i.e.,
$w_{\rm q} \in [-1,\,0]$.

Here we introduce a modified model of quintessence when the dark
energy and the corresponding pressure obeys the following equation
of state:
\begin{equation}
p_{\rm DE} = -w(\ve_{\rm DE}-\ve_{\rm cr}), \label{quintnew}
\end{equation}
where the constant $w \in [0,\,1)$. Here $\ve_{\rm cr}$ some
critical energy density.  Setting $\ve_{\rm cr} = 0$ one obtains
ordinary quintessence. It is well known that as the Universe
expands the (dark) energy density decreases. As a result, being a
linear negative function of energy density, the corresponding
pressure begins to increase. In case of an ordinary quintessence
the pressure is always negative, but for a modified quintessence
as soon as $\ve_{\rm q}$ becomes less than the critical one, the
pressure becomes positive. In Fig. \ref{presqom} we illustrate the
evolution of pressure corresponding to ordinary and modified
quintessence with the expansion of the Universe.

\myfig{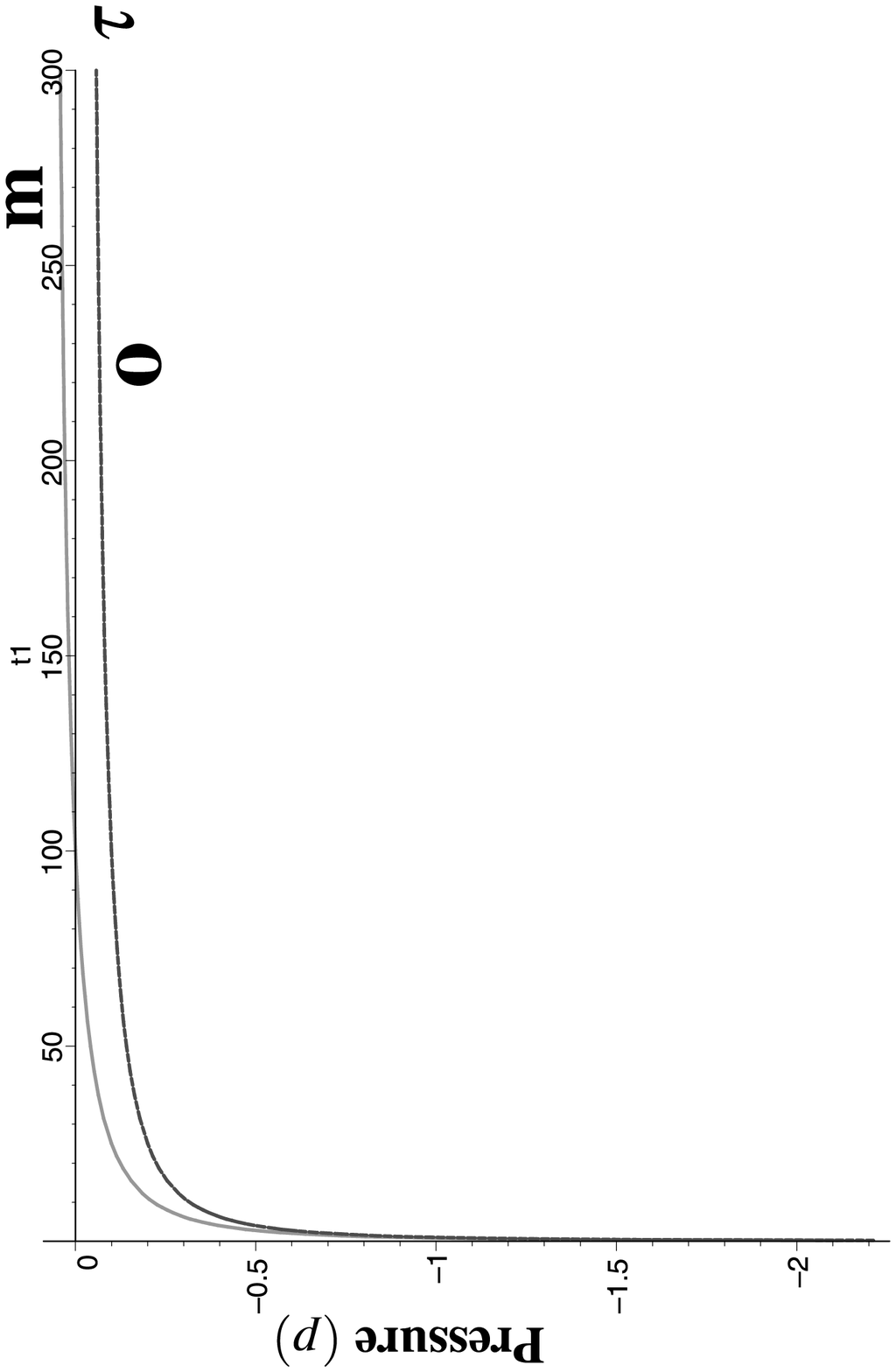}{0.33}{Evolution of the pressure with the expansion
of the BI Universe when it is
 filled with perfect fluid and dark
energy obeying ordinary and modified quintessence given by
\eqref{quint} and \eqref{quintnew}, respectively. Here the letters
{\bf "o"} and {\bf "m"} stand for ordinary and modified,
respectively.}

In account of \eqref{quintnew} from \eqref{deve} one finds the
following relation between $\ve_{\rm DE}$ and $\tau$ :
\begin{equation}
\ve_{\rm DE} = \frac{1}{1 - w}\Bigl[\frac{\ve_{1}}{\tau^{1- w}} -
w\ve_{\rm cr}\Bigr],
\end{equation}
with $\ve_{1}$ being some integration constant.

\subsection{Exact and numerical solutions}

As soon as the right hand side of the Eq. \eqref{dtau1} is
defined, we can study this equation in details. First we write the
solution to the equation in question in quadrature which will be
followed by numerical results.

Inserting $\ve_{\rm pf}$ and $\ve_{\rm q}$ into \eqref{quad} one
now finds
\begin{equation}
\int \frac{d\tau}{\sqrt{C_1 + 3 \kappa\bigl(\ve_0 \tau^{(1
-\zeta)} + [\ve_{1}/(1-w)]\tau^{(1 +w)} - [w\ve_{\rm cr}/(1-w)]
\tau^2\bigr)}} = t + t_0. \label{quadq}
\end{equation}
Here $t_0$ is a constant of integration that can be taken to be
trivial. As one sees, the positivity of the radical imposes some
restriction on the maximum value of $\tau$, i.e., in this case the
model allows oscillatory mode of expansion. It means the dark
energy initially acts as a repulsive force resulting in
accelerated expansion of the Universe. But as soon as $\ve_{\rm
DE}$ becomes less than $\ve_{\rm cr}$ the corresponding pressure
changes its direction and as a result the Universe begins to
contract. In order to give a complete picture, beside the system
with a modified quintessence we study the system with an ordinary
one as well. It should be noted that the critical energy density
$\ve_{\rm DE}$ should be very small. Here for simplicity we set
$\ve_{\rm cr} = 0.01$.

\myfigs{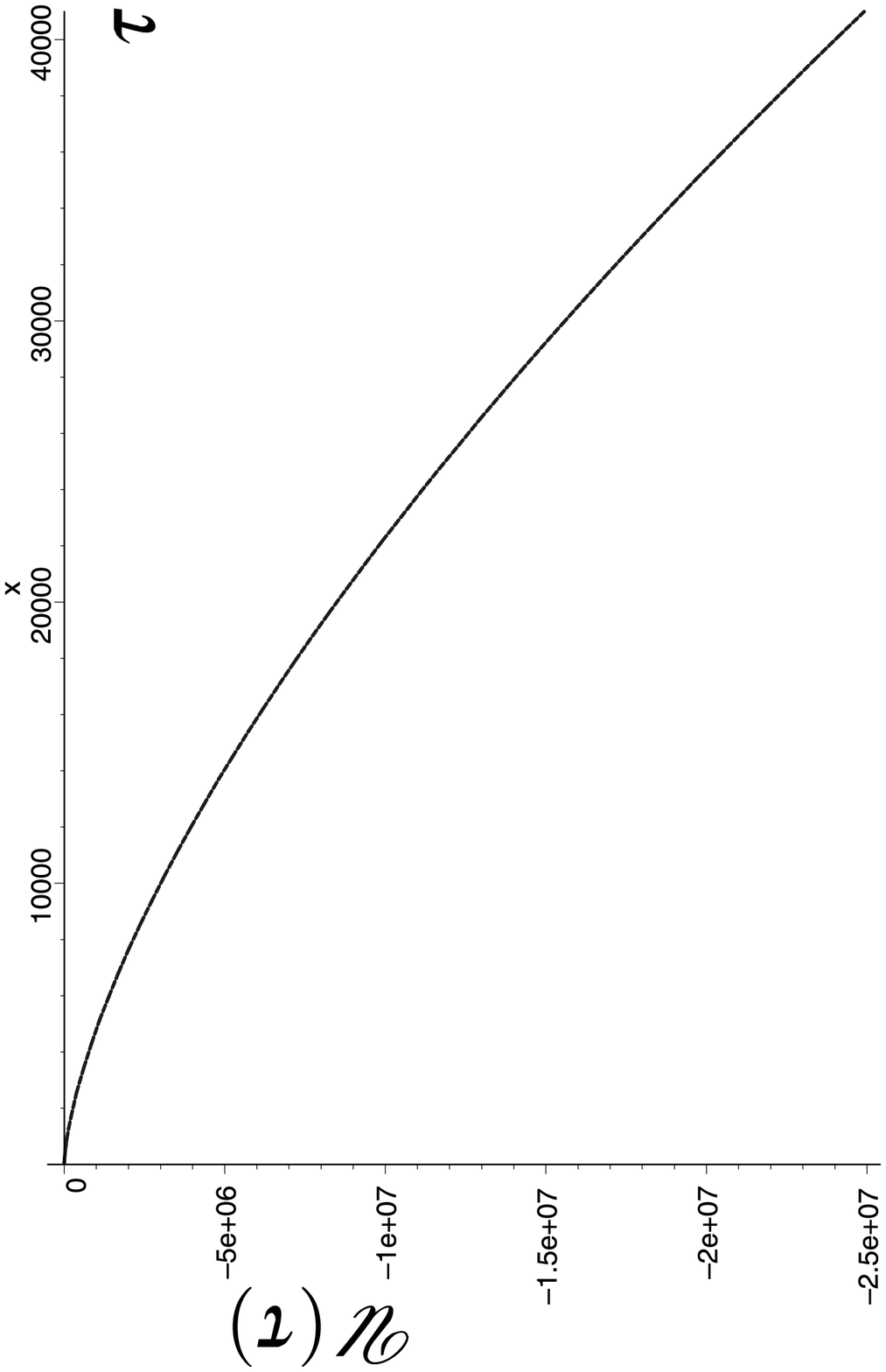}{0.30}{View of the potential when the BI Universe is
filled with a binary mixture of perfect fluid and a quintessence
given by \eqref{quint}.}{0.43}{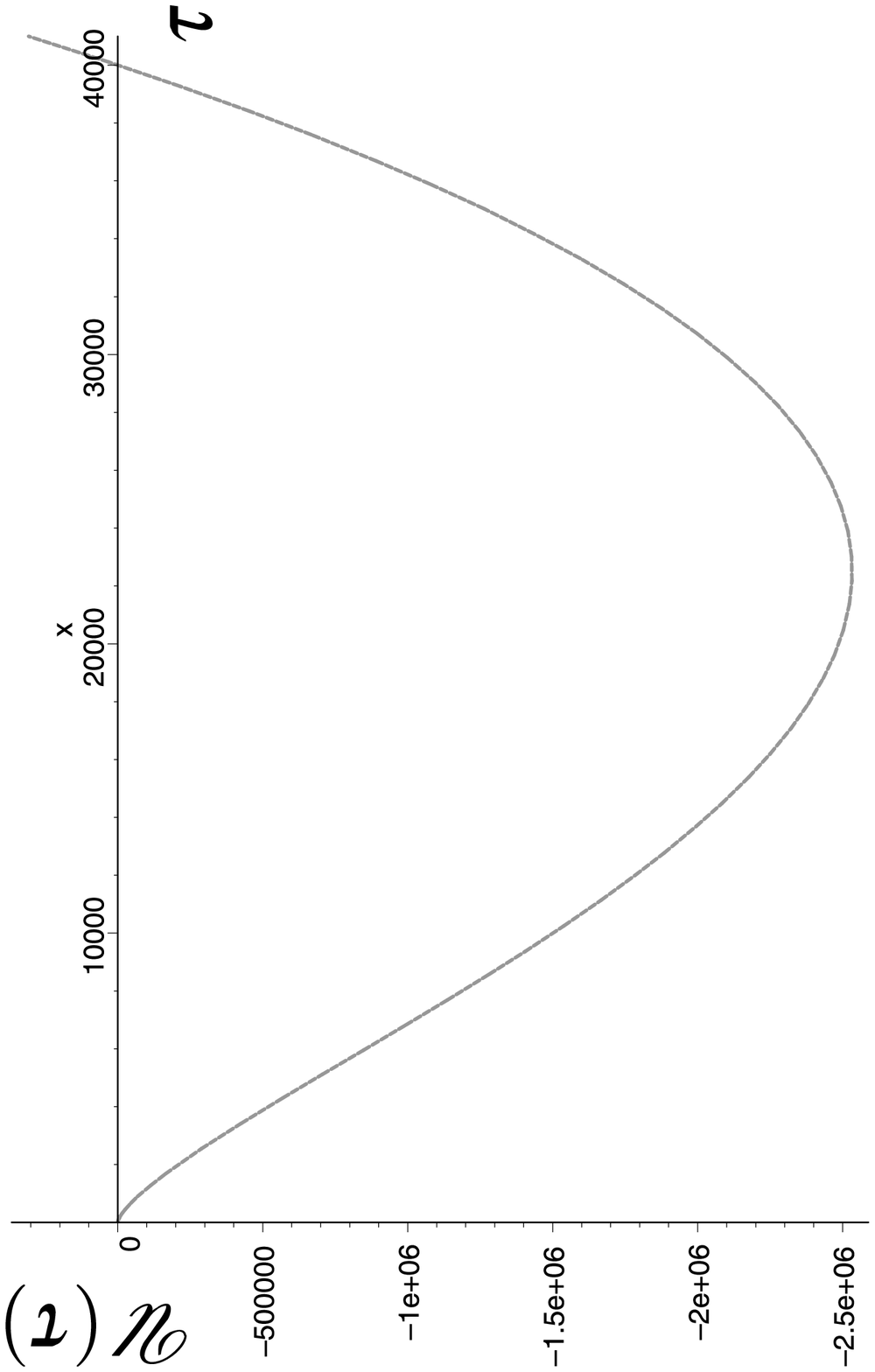}{0.30}{View of the potential
when the BI Universe is filled with a binary mixture of perfect
fluid and a modified quintessence-like dark energy given by
\eqref{quintnew}.}{0.43}

In Figs. \ref{pot00} and \ref{poten} potentials corresponding to
an ordinary and modified quintessence are given. As one sees, the
usual quintessence does not allow oscillatory mode of expansion,
Universe in this case expands endlessly. A corresponding expansion
of the Universe is given in Fig. \ref{tauq}. It should be
emphasized that the initially anisotropic Universe in this case
evolves into an isotropic FRW one. As oppose to the ordinary
quintessence a modified quintessence imposes some restriction on
the maximum value of $\tau$. As a result the Universe initially
expands, but after reaching some maximum it begins to contract.
Depending on the choice of the constant, which can be viewed as an
energy level, it either shrinks into a point [cf. Fig.
\ref{taunonpc1}] thus giving rise to space-time singularity, or
begins to expand again after reaching some non-zero minimum, i.e.,
the Universe in this case experiences the oscillatory mode of
expansion [cf. Fig. \ref{tauqmod}]. Note that to each ${\cal E}$
corresponds a particular pair $(\tau_{\rm min}, \tau_{\rm max})$.

\myfigs{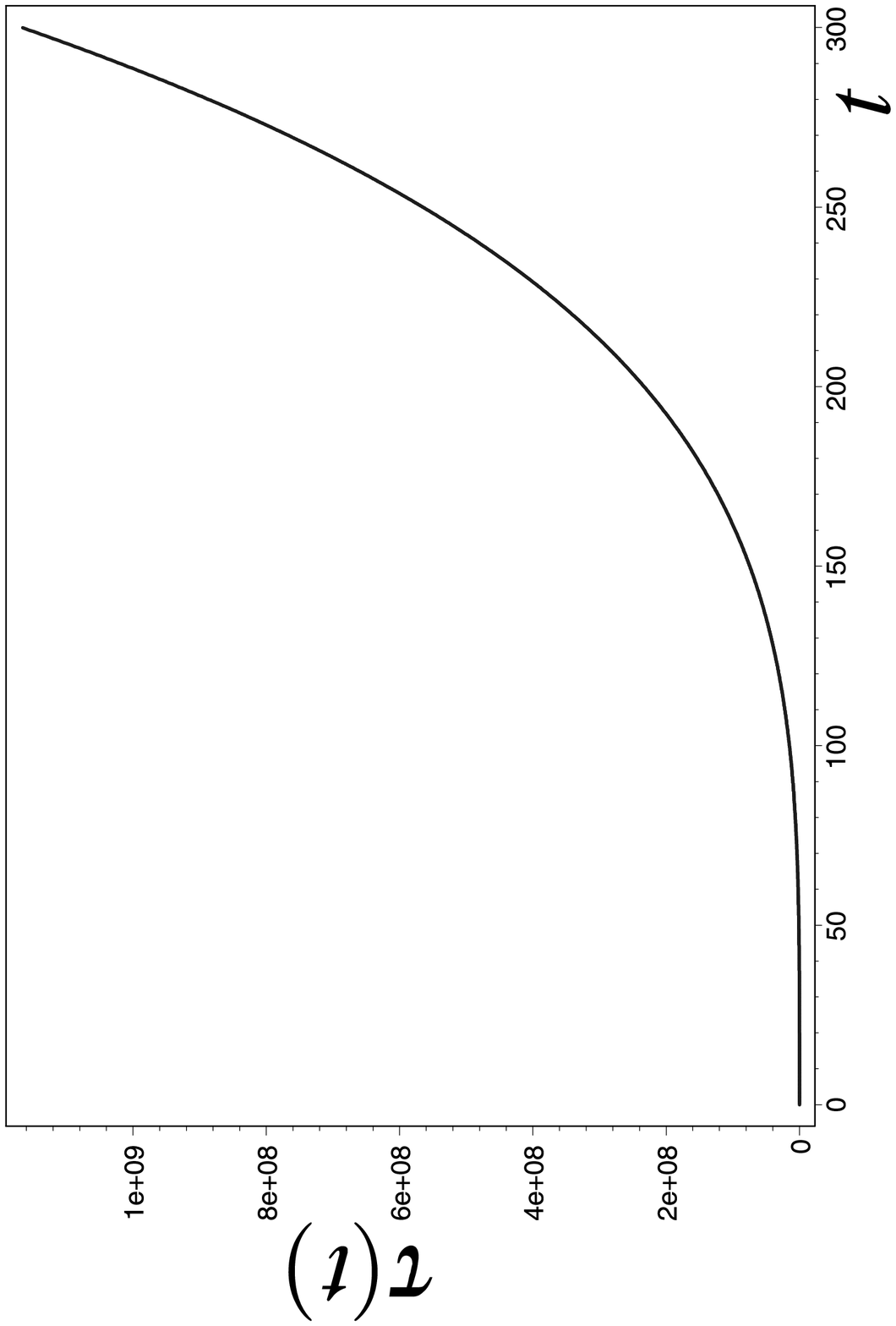}{0.30}{Evolution of the BI Universe is filled with a
binary mixture of perfect fluid and a quintessence. For
simplicity, as a perfect fluid we consider only radiation.}
{0.43}{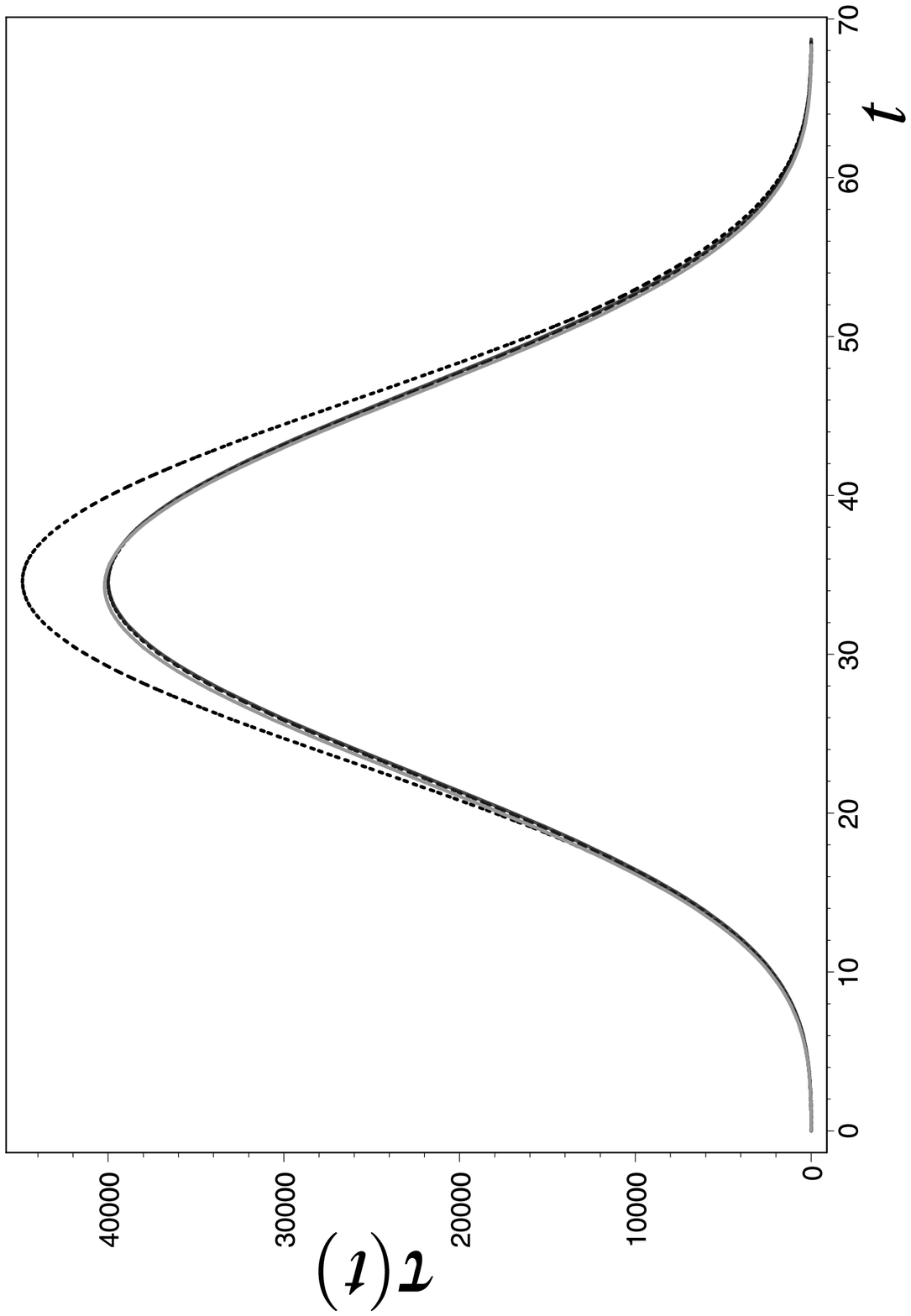}{0.30}{Evolution of the BI Universe with a
modified quintessence. Choosing ${\cal E} \ge 0$ (here we set
${\cal E} = 1$ one obtains nonperiodic picture of evolution. As
one sees, different choice of $\zeta$ gives rise to different
amplitude, but the overall character of the solution remains
unaltered.}{0.43}

\myfigs{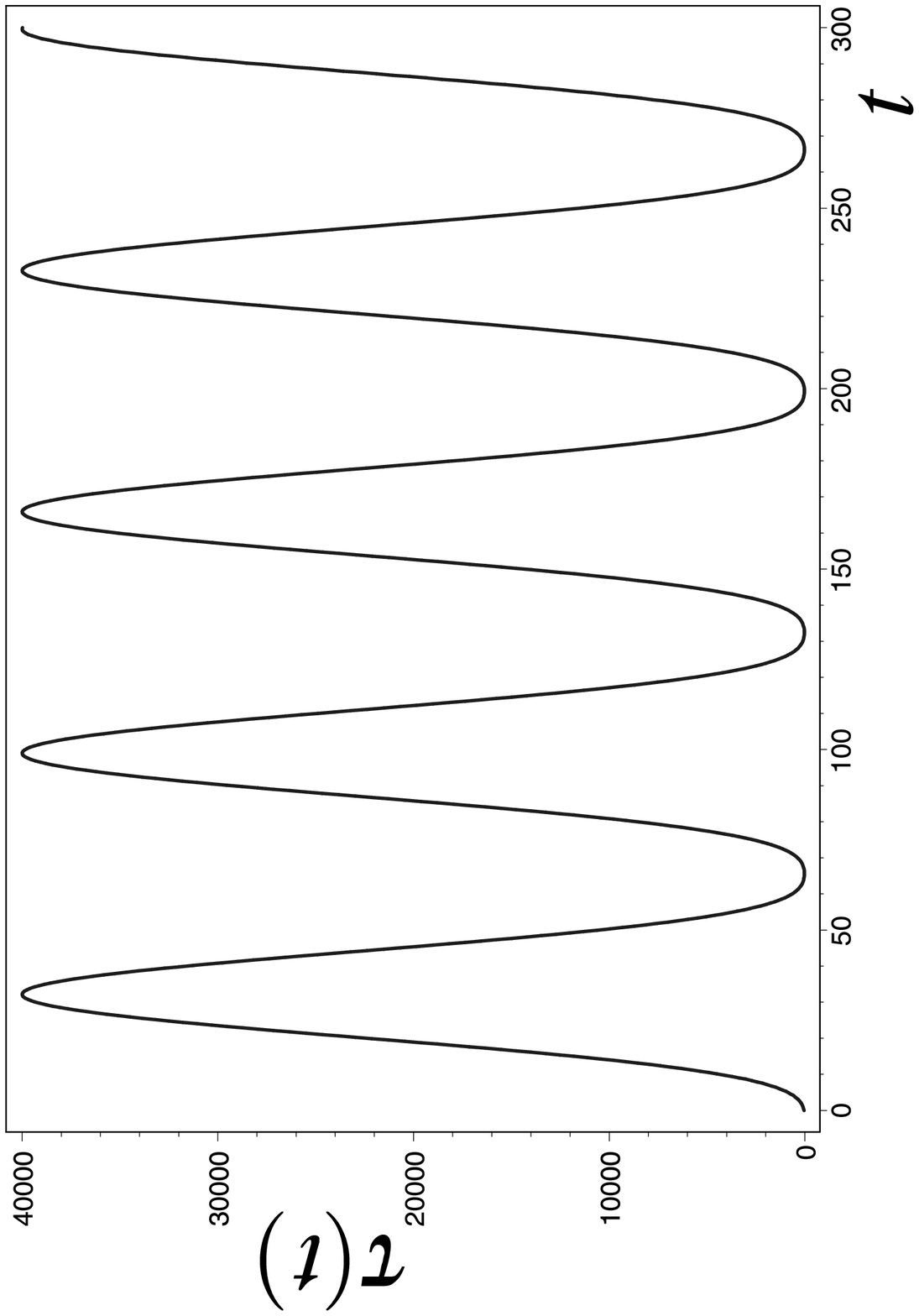}{0.30}{For ${\cal E} <  0$  BI Universe filled
with a binary mixture of perfect fluid and quintessence-like dark
energy admits oscillatory mode of expansion. Here we set ${\cal E}
= -1000$ with $\tau_0 = 50$.}{0.43}{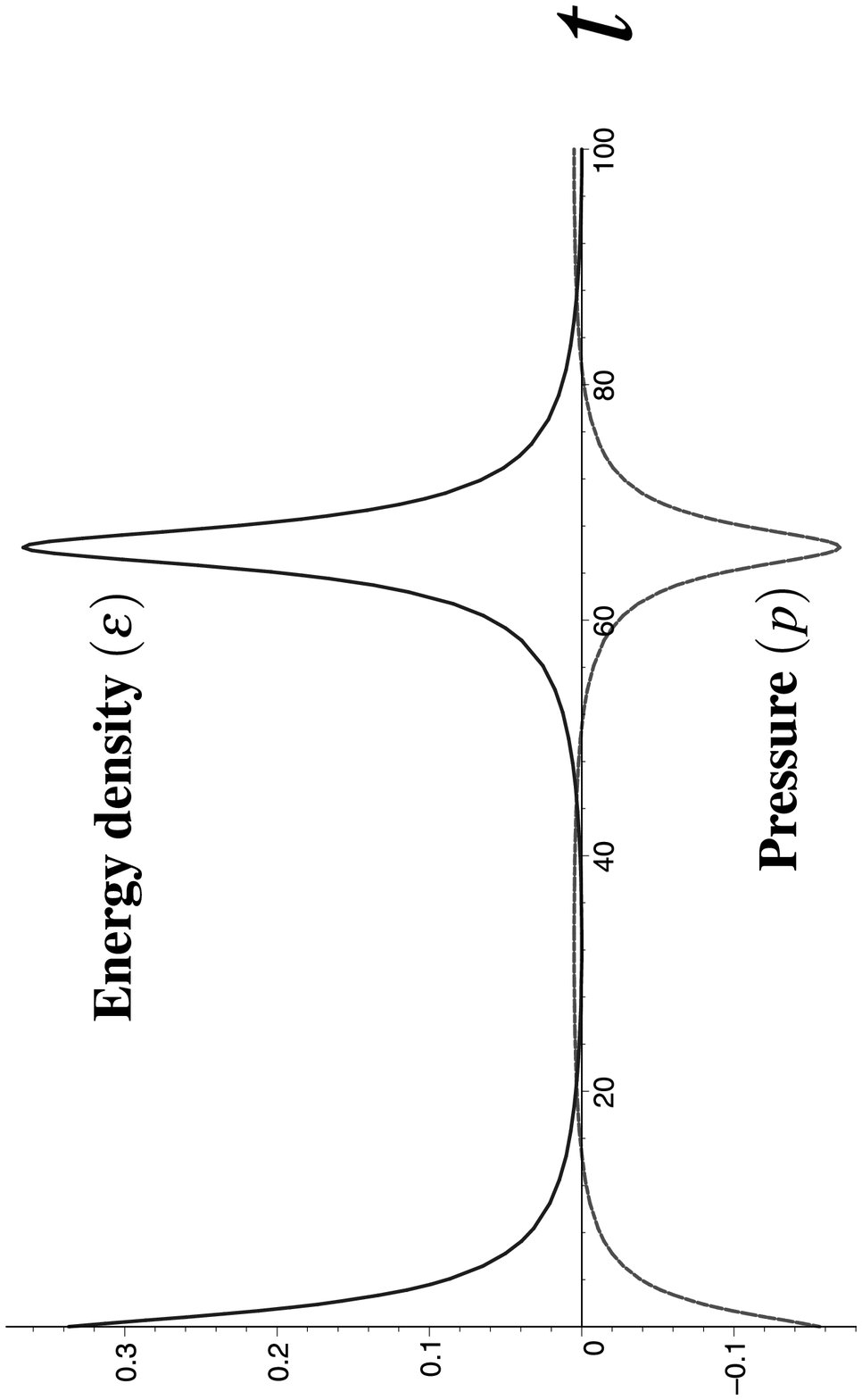}{0.30}{View of energy
density and pressure when BI Universe experiences oscillation as
in Fig. \ref{tauqmod}.}{0.43}

In Fig. \ref{edpr} we illustrate the evolution of energy density
and pressure when the Universe is filled with a binary mixture of
perfect fluid given by a radiation and a modified quintessence. As
one sees, the initial density is large enough and the pressure is
initially negative, with the energy density being less than
$\ve_{\rm cr}$, pressure becomes positive and the Universe begins
to contract. As a result the energy density begins to increase and
pressure again becomes negative. This results in the oscillatory
mode of expansion.

\myfig{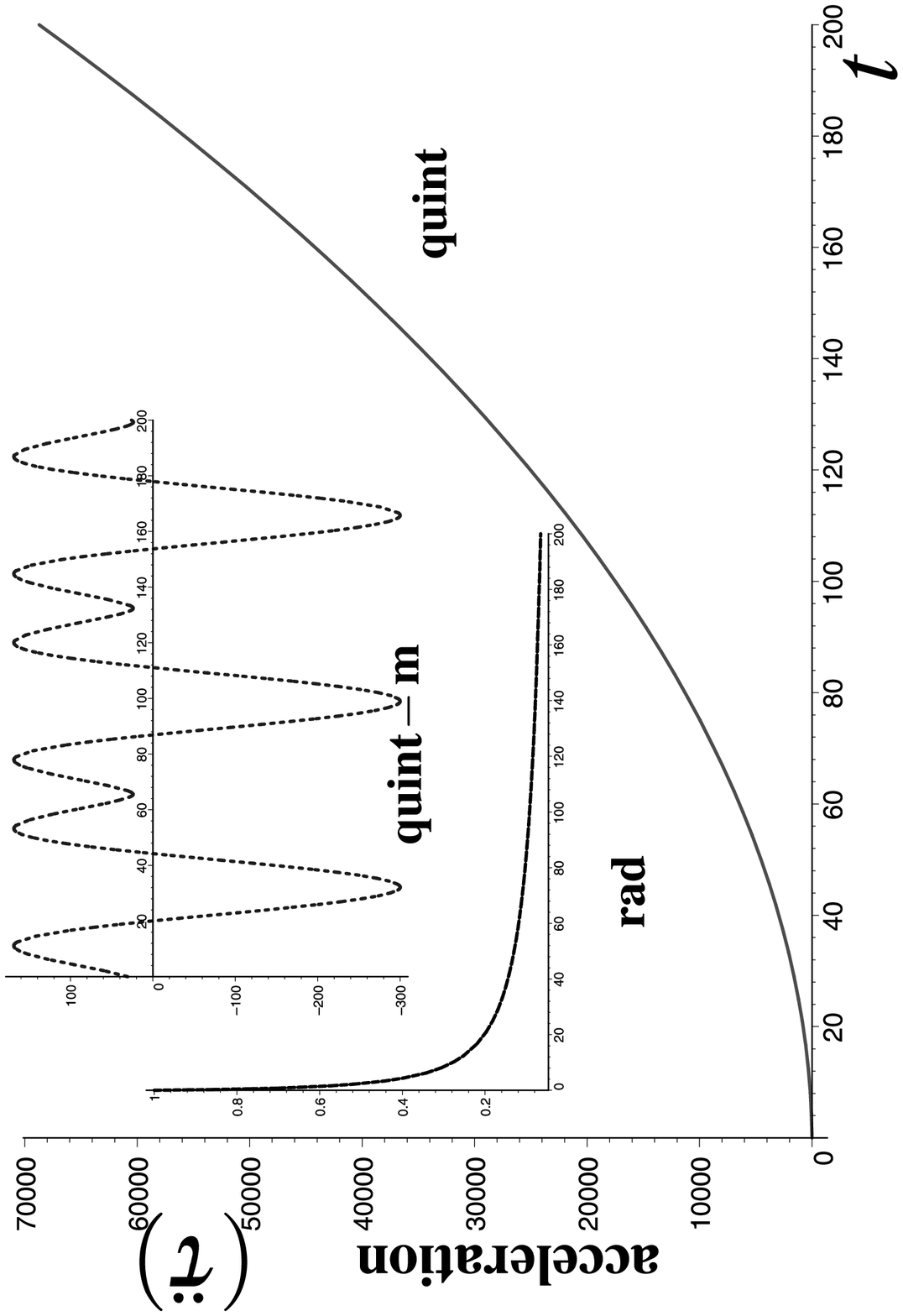}{0.33}{View of the acceleration for different
source fields. Here "rad", "quint" and "quint-m" stand for
radiation, a mixture of radiation and an ordinary quintessence and
a mixture of radiation and modified quintessence, respectively.}

Finally in Fig. \ref{acceleration} we plot the graphic of
acceleration verses time. As one sees, if the Universe is filled
with perfect fluid only, the process of evolution decelerates with
time, while the inclusion of an ordinary quintessence results in
an accelerated evolution. If the dark energy is given by a
modified quintessence the accelerated expansion is followed by a
decelerated one.

\section{Conclusion}

A self-consistent system of BI gravitational field filled with a
perfect fluid and a dark energy given by a modified version of
quintessence has been considered. The exact solutions to the
corresponding field equations are obtained. The inclusion of the
dark energy into the system gives rise to an accelerated expansion
of the model. As a result the initial anisotropy of the model
quickly dies away. The modification of the quintessence results in
appearing some upper bound of the volume scale, i.e., in this case
the Universe is spatially finite. Depending on the choice of the
integration constant ${\cal E}$, which can be viewed as an energy
level, the Universe is either close with a space-time singularity,
or an open one which is oscillatory, regular and infinite in time.

\newcommand{\hnl}{\htmladdnormallink}

\end{document}